\documentclass{iau} 
\usepackage{graphicx}
\usepackage{caption}
\usepackage{subcaption}

\title[source regions and $^3$He-rich impulsive solar energetic particles events] 
{Temperature of source regions of $^3$He-rich impulsive solar energetic particle events}

\author[Chen, Bu{\v c}{\'{\i}}k and Kim]   
{N.-H. Chen$^1$,
R. Bu{\v c}{\'{\i}}k$^{2,3}$
 \and R.-S. Kim$^{1,4}$}

\affiliation{$^1$Korea Astronomy and Space Science Institute, Daejeon, Republic of Korea \\ email: {\tt nhchen@kasi.re.kr} \\[\affilskip]
$^2$Institut f$\ddot{u}$r Astrophysik, Georg-August-Universit$\ddot{a}$t G$\ddot{o}$ttingen, G$\ddot{o}$ttingen, Germany \\
$^{3}$ Max-Planck-Institut f$\ddot{u}$r   Sonnensystemforschung, G$\ddot{o}$ttingen, Germany \\
$^{4}$ University of Science and Technology, Daejeon, Republic of Korea \\}
\pubyear{2017}
\volume{335}  
\jname{Space Weather of the Heliosphere: Processes and Forecasts}
\editors{Claire Foullon, \& Olga Malandraki, eds.}

\begin{document} 
\maketitle

\begin{abstract}
Impulsive solar energetic particle (SEP) events originate from the energy dissipation process in small solar flares. Anomalous abundances in impulsive SEP events provide an evidence on unique, yet unclear, acceleration mechanism. The pattern of heavy-ion enhancements indicates that the temperature of the source plasma that is accelerated is low and not flare-like. We examine the solar source of the $^3$He-rich SEP event of 2012 November 20 using Solar Dynamics Observatory (SDO)/ Atmospheric Imaging Assembly (AIA) images and investigate its thermal variation. The examined event is associated with recurrent coronal jets. The Differential Emission Measure (DEM) analysis is applied to study the temperature evolution/distribution of the source regions. Preliminary results show that the temperature of the associated solar source is ranged between 1.2-3.1 MK.
\keywords{Sun: UV radiation, particle emission, corona}
\end{abstract}

\section{Introduction}
{Impulsive solar energetic particles (SEP) events, discovered more than 40 years ago, are understood to come from the energy dissipation process in solar flares. One remarkable feature of impulsive SEP events is the enhanced abundance of $^3$He isotopes and other heavier ions compared to corona. The in-situ observations of ion composition may provide important information on a possible nature of the sources as well as a clue on mechanism of particle acceleration on the sun. \\
\indent It has been known from earilier observations that element abundance enhancement is an increasing function of mass-to-charge ratio (A/Q) of ions in impulsive SEP events \cite[(e.g., Reames et al. 2014; 2015a,b)]{reames14}. An examination of theoretical A/Q profiles for various ion species at equilibrium temperature T indicates that any subtle variations in the temperature of source plasma could play a significant role in the properties of SEP events detected later. A narrow temperature range of 2-4 MK in the source region has been inferred to be responsible for the observed element abundances \cite[(Reames et al. 2014)]{reames14}, though some studies \cite[(Mason et al. 2002, 2016)]{mason16} have proposed even lower temperatures ($<$ 1.5 MK). A direct measurement of charge states for low energy ions is consistent with the ambient plasma temperature \cite[(DiFabio et al. 2008)]{DiFabio08}. The higher charge states for high-energy ions are suggested to be due to charge stripping after acceleration. Thus the temperature of the source region in impulsive SEP events is uncertain as well as whether the temperature solely determines which element is enhanced during the acceleration. \\
\indent With improved remote sensing observations (such as SDO/AIA) now, we are able to derive the detailed temperature distribution in active regions and thus to investigate the conditions for the enhancement mechanism in $^3$He-rich impulsive SEP events. In this work, we  report for the first time the derived temperature distribution of the source region in one $^3$He-rich impulsive SEP event. It is shown in section 2. The discussion and summary is given in section 3.}
\begin{figure}
\graphicspath{ {C:/Users/Martina/Desktop/work/KASI_work/conference/IAU_SW/proceeding/fig/} }
\centering
\begin{subfigure}{.53\textwidth}
  \centering
  \includegraphics[width=1\linewidth]{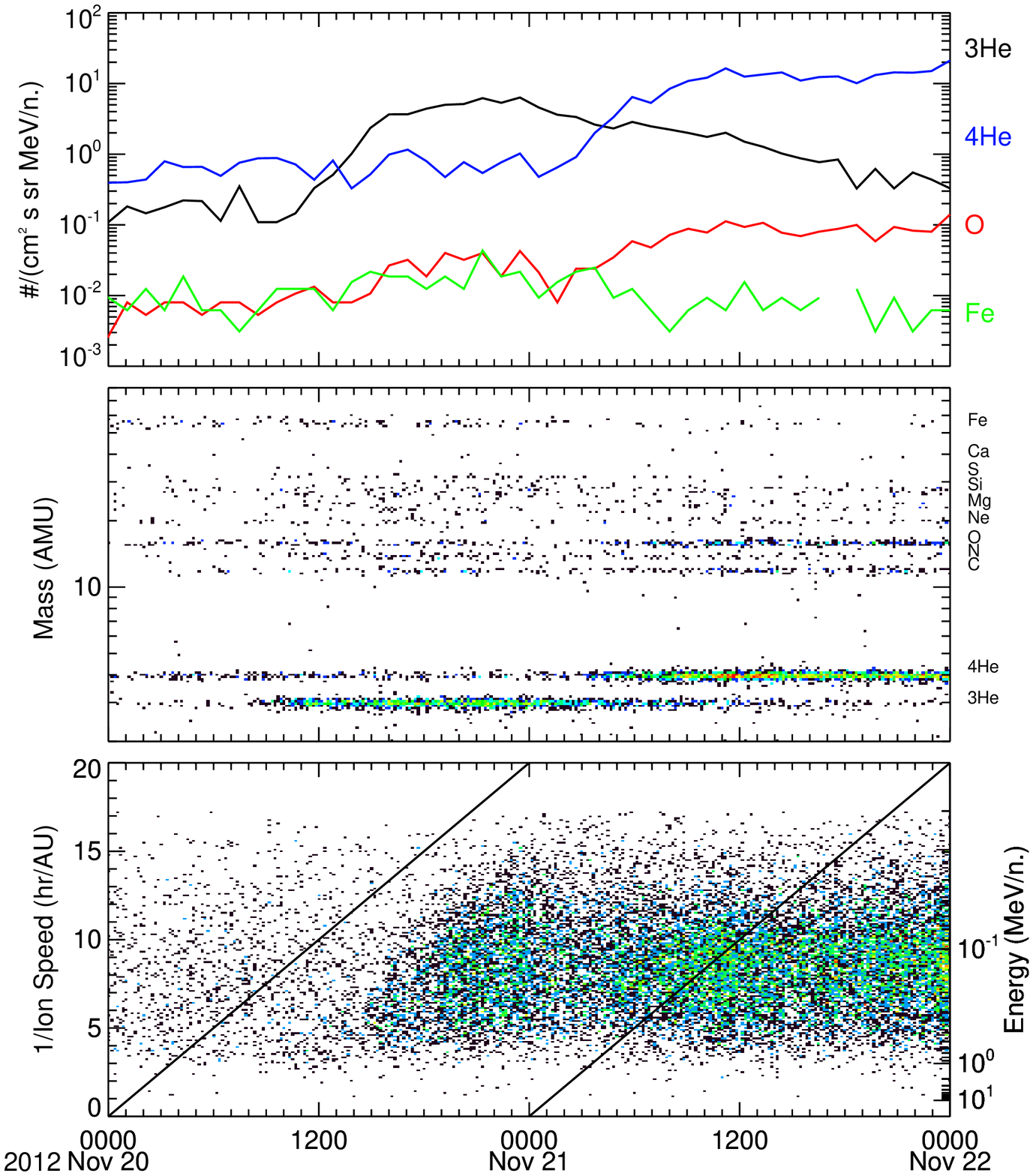}
  \caption{}
  \label{fig:sub1}
\end{subfigure}%
\begin{subfigure}{.48\textwidth}
  \centering
  \includegraphics[width=1\linewidth, trim=3cm 13.2cm 9cm 5cm, clip=true ]{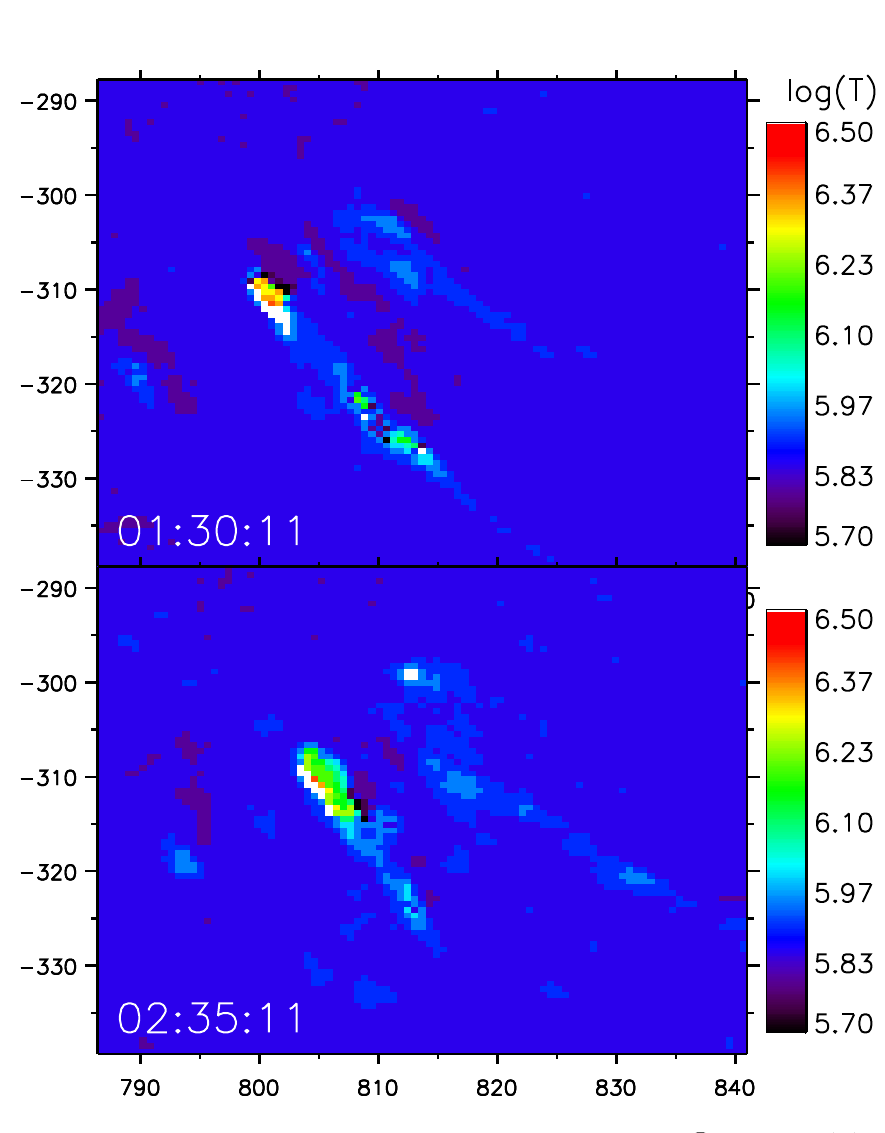}
  \caption{ }
  \label{fig:sub2}
\end{subfigure}
\caption{(a) $^3$He-rich SEP event. The panels from top to bottom are: 1 hr Intensity profile of $^4$He, $^3$He, O and Fe in energy range 0.23 - 0.32 MeV/nucleon; Mass spectrogram in energy range 0.4-10 MeV/nucleon and Inverted ion velocity-time spectrogram in mass range 10-70 amu.\\
(b)Peak temperature map of the event source. The colorbar aside gives the magnitude of temperature. The selected subfigures are around the onset time of the accompanied type III burst. }
\label{fig:test}
\end{figure}
\section{Observations and data analysis}
{ \indent The event under study was observed on 2012 November 20 (see figure 1a). This prominent event started around 10:00 UT at 0.23-0.32 MeV/nucleon with a large $^3$He enrichment and was intervened at the beginning of November 21 by a large-gradual SEP event with a normal coronal composition.
Its source region has been identified as active region NOAA 11617 based on our identification method, including the connection analysis via the interplanetary and coronal field model and spatial-temporal coincidence on the EUV multi-wavelengths observation with the type III radio bursts \cite[(Chen et al. 2015)]{chen15}. The associated solar eruption in the source region consists of two jets, accompanied by strong type III radio bursts with the onset at 01:29 and 02:32 UT, showing a deep drift below 0.1 MHz (see Figure 2 in \cite[Chen et al. (2015)]{chen15}). Besides, the event is characterized by an enormously enhanced S/O ratio, $\sim$5 times higher than typical $^3$He-rich events S/O ratio \cite[(Mason et al. 2016)]{mason16}. \\
\indent To examine the temperature distribution at the source region, we utilized the Differential Emission Measure (DEM) analysis using six EUV channels (94, 131, 171, 93, 211, 335 {\AA}) of SDO/AIA observations \cite[(Aschwanden et al. 2013)]{aschwanden13}. A Gaussian function is used to estimate the peak value of the emission measure and temperature that produce the temperature map (see Figure 1b). We calculated the temperature of two jets at the onset of type III bursts which may be close to the time of particle acceleration/injection. Along the jet, the distribution of temperature is patch-like, inhomogeneous and some parts are saturated (indicated by white color). Generally, the footpoints (the root of jets) reveal a temperature value  higher than the one in jet's spire. As far as the standard 2D reconnection model is concerned, the footpoints represent the closed loops where most of the thermal energy is confined and the spire indicates the open field regions where all the energy would be transported outward. The estimated temperature of source plasma, including footpoint and spire, is relatively smaller than the theoretical value. Note that \cite[Mason et al. (2016)]{mason16} inferred that the temperature for S-rich events (including the one examined) is cooler with lower boundary at 0.4 MK. It appears that ions in the examined event were accelerated along the cooler jet spire. }
\section{Summary}
{The solar source of the $^3$He-rich SEP event has been studied by analyzing high cadence SDO/AIA observations. The source region produces several jets with associated type III/VI radio bursts. The temperature of associated source eruptions is derived from DEM analysis based on EUV observations. In this events, the jet's spire is cooler than the footpoint regions. The estimated temperature of both jets, including footpoint and spire, ranges from 1.25- 3.1 MK which is smaller than the standard value of 2-4 MK inferred from elemental abundance observations \cite[(Reames et al. 2014)]{reames14}. In future works, we plan to characterize the temperature of the source regions in other (eg. $^3$He-rich, Fe-rich) SEP events.}

\section*{Acknowledgement}
The work of R. Bu{\v c}{\'{\i}}k is supported by the Deutsche Forschungsgemeinschaft under grant BU 3115/2-1.


\end{document}